\documentclass[prb,twocolumn,superscriptaddress,preprintnumbers,amsmath,amssymb,floatfix]{revtex4-2}

\usepackage{graphicx}
\usepackage{mathrsfs}
\usepackage{color}

\begin{document}

%=======================================================================================
\title{Quantum shockwave at the quasi-relativistic resonance}

\author{Marinko Jablan}
\email{mjablan@phy.hr}
\affiliation{Department of Physics, Faculty of Science, University of Zagreb, 10000 Zagreb, Croatia}

\date{\today}

\begin{abstract}
Shockwaves are violent nonlinear distortions of wave motion which have been reported in fluid waves and electromagnetic waves, while here we reveal that a shockwave can occur even in the quantum wave function of a single particle. Specifically, we analyze the electromagnetic field travelling on resonance with the limiting velocity of the quasi-relativistic particle, and reveal that a strong longitudinal field can rip up a quantum shockwave singularity in the particle wave function, leading to energy dissipation. Moreover, we show that this effect is particularly strong for quasi-relativistic (Dirac) electrons in graphene for two reasons. On one hand, we obtain the single-particle shockwave at very small fields due to the small effective electron mass, and on the other hand we can obtain large dissipation in the many-particle shockwave by using the large electron density in graphene. While the single-particle shockwave can be most easily observed by locating the shockwave singularity with a high-resolution microscopy, the many-particle dissipation would be simply observed as a rapid decay of our resonant electromagnetic field.
\end{abstract}

\maketitle

%\narrowtext
%\newpage

Einstein theory of relativity postulates the limiting velocity of signal propagation which no massive particle can reach. Only massless particles of light can propagate at this velocity, however since they don't have a rest frame, the relativistic symmetry enables the light with only two transverse polarizations while the longitudinal polarization is forbidden \cite{QED}.
On the other hand, one can have quasi-relativistic particles \cite{Dirac14} which have low-energy behavior reminiscent of the  relativistic particles with the quasi-limiting velocity of propagation (which is of course now less than the velocity of light). This offers us an opportunity to explore the effects of the longitudinal electromagnetic (EM) fields on these quasi-relativistic particles. In fact, it was recently shown that longitudinal field propagating at velocity close to resonance with this quasi-limiting velocity results in the large nonlinear response which apparently diverges on resonance \cite{Jablan20}. Here we show that strong field exactly on resonance results in the quantum shockwave accompanied by energy dissipation, while the particle gets localized at the quantum wave function singularity. 
 
To give some specific sense of the scale we focus on the low energy electrons in two-dimensional (2D) graphene layer described by the relativistic Dirac Hamiltonian operator \cite{Dirac14}:
\begin{equation}
\hat{H}=c\sigma_x\hat{p}_x+c\sigma_y\hat{p}_y+mc^2\sigma_z,
\end{equation}
where $\sigma_{x,y,z}$ are Pauli spin matrices, $\hat{p}_{x,y}=-ih\partial_{x,y}$ are components of 2D momentum operator, $c=10^6$ m/s is the quasi-limiting velocity, while the effective mass (e.g. due to spin-orbit coupling) is typically very small (with the rest energy $mc^2\approx 10\ \mu\mbox{eV}$). It is then straightforward to find the free particle eigenfunctions: 
\begin{equation}
\psi_0({\bf r},t)=e^{\frac{i}{\hbar}({\bf p}\cdot{\bf r}-Et)}\phi_0=
e^{\frac{i}{\hbar}({\bf p}\cdot{\bf r}-Et)}
\left( {\begin{array}{c}
1 \\
\frac{c(p_x+ip_y)}{E+mc^2} \\
\end{array}} \right),
\label{eigenfunction}
\end{equation}
where ${\bf r}=(x,y)$ is a 2D position vector, ${\bf p}=(p_x,p_y)$ is the momentum eigenvalue, while the energy eigenvalue satisfies the relativistic-like relation:
\begin{equation}
E^2=c^2p_x^2+c^2p_y^2+m^2c^4.
\end{equation}

Let us now introduce the longitudinal EM field propagating in $x$-direction at this quasi-limiting velocity $c$, described by the vector potential: ${\bf A}=(A(u),0)$, where $u=ct-x$. Electron wave function $\psi$ is then determined by the Dirac equation:
\begin{equation}
\left[c\sigma_x(\hat{p}_x-eA(u))+c\sigma_y\hat{p}_y+mc^2\sigma_z-i\hbar\partial_t\right]\psi=0,
\label{Dirac}
\end{equation}
where $e=-|e|$ is the electron charge. However, before discussing the full quantum solution, it is instructive to study first briefly the quasi-classical solution: $\psi^{qc}=ae^{iS/\hbar}$, where $a$ is a slowly varying amplitude and $S/\hbar\gg1$ is a large phase \cite{QM}. It is then easy to show that $S$ is just the classical action determined by the Hamilton-Jacobi equation \cite{Jablan20}:
\begin{equation}
(\partial_tS)^2=c^2(\partial_xS-eA(u))^2+c^2(\partial_yS)^2 +m^2c^4,
\end{equation}
which can be easily solved with the following ansatz \cite{Jablan20}:
\begin{equation}
S({\bf r},t)={\bf p}\cdot{\bf r}-Et+s(u),
\label{S}
\end{equation}
to obtain
\begin{equation}
s(u)=\int_{-\infty}^u du\frac{c}{2}\frac{2p_xeA-e^2A^2}{E-cp_x+ceA},
\label{s}
\end{equation}
where we have assumed that the field is turned off in the distant past: $A(u=-\infty)=0$, so that we start with a free particle state. Note however that we encounter the problem at the singularity of the integrand:
\begin{equation}
E-cp_x+ceA_s=0.
\label{singularity}
\end{equation}
In fact, since $\dot s=ds/du$ diverges at the singularity, one can easily show that both the classical energy: $E^c=-\partial_tS=E-c\dot s$, and momentum: $p_x^c=\partial_xS=p_x-\dot s$, also diverge at this singularity. This is a consequence of the fact that the classical velocity approaches the quasi-limiting velocity at this singularity: $v_x^c=\partial E^c/\partial p_x^c \rightarrow c$, which means that there is a resonant transfer of energy between the particle and the field moving at the velocity $c$. Particularly, for $E>0$ (i.e. $E<0$) the field gives (takes) energy to (from) the particle, respectively.
 
Moreover, note that we get a multivalued solution in $S$ as we are apparently free to avoid the singularity in the integral (\ref{s}) by any which way in the complex $u$-plane. To distinguish an actual physical solution, we need to include (at least an infinitesimal) coupling to the environment, which we take into account by the Landau rule: $E\rightarrow E+i\eta$, where we can put $\eta\rightarrow 0^+$ only at the very end of calculation \cite{QM}. If the reader is uncertain of the correct sign of $\eta$ (as was the writer) it is best to simply follow the Landau advice and fix the sign so that one gets a physical result in the end (this being the correct sign of the energy dissipation in our case). This type of argument (invoking the second law of thermodynamics) is very useful in dealing with shockwaves \cite{FM, ECM}, however we also give another argument below to show that $\eta>0$. Then, via the relation:
\begin{equation}
\frac{1}{x+i0^+}=\mathscr{P}\frac{1}{x}-i\pi\delta(x),
\label{pole}
\end{equation}
(where $\mathscr{P}$ denotes the principal value) we obtain imaginary part od the action:
\begin{equation}
\Im S(u)=\frac{E^2-c^2p_x^2}{c^2}\frac{\pi}{2}\sum_{n=1}^N\frac{\theta(u-u_n)}{|e\dot A_n|},
\end{equation}
where $\theta(u)$ is the unit step function, and $u_n$ is one of (say $N$) possible solutions of the Eq. (\ref{singularity}). Note that imaginary action doesn't make any sense classically, but in the quasi-classical case represents jump in the particle probability density $|\psi^{qc}|^2\sim e^{-2\Im S/\hbar}$. We will in fact show that we get quantum wave function singularity in the form of a shockwave accompanied by energy dissipation.

\begin{figure}
\centerline{\mbox{\includegraphics[width=0.5\textwidth]{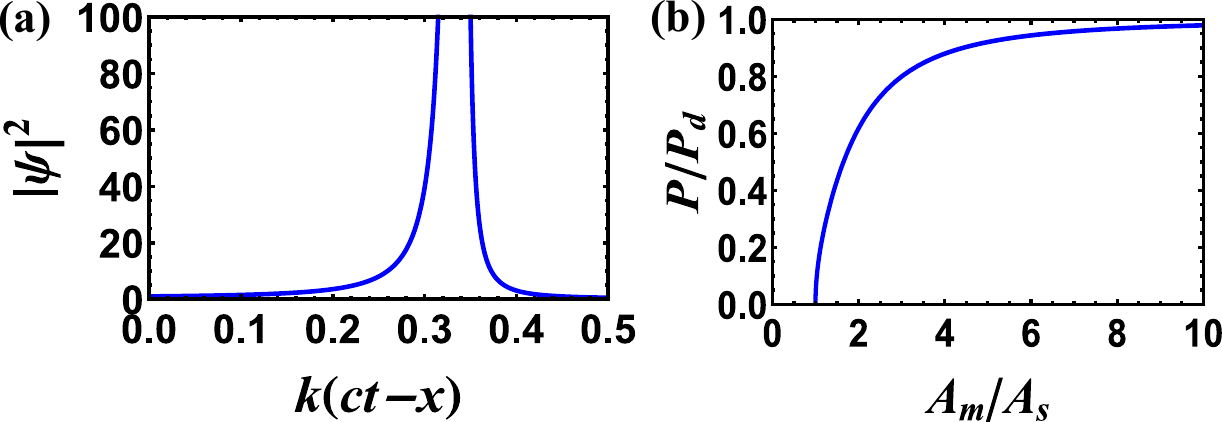}}}
\caption{
(a) Quantum shockwave singularity of a quasi-relativistic particle, and (b) power dissipation in the potential: $A_x(x,t)=A_m\sin k(ct-x)$.
}
\label{fig1}
\end{figure}

One can reduce the full quantum case to the simple interaction between quasi-classical states as was done in the off resonant case in the reference \cite{Jablan20}. However, it is much more convenient to follow the original Volkov approach \cite{QED} to deal with the resonant case here. Therefore, we first transform the first order Eq. (\ref{Dirac}) by the action of the operator:
 \begin{equation}
c\sigma_x(\hat{p}_x-eA(u))+c\sigma_y\hat{p}_y+mc^2\sigma_z+i\hbar\partial_t,
\end{equation}
into the second order Eq.:
\begin{equation}
\begin{split}
&[c^2(-i\hbar\partial_x-eA)^2+c^2(-i\hbar\partial_y)^2+m^2c^4+\hbar^2\partial_t^2 \\
&-i\hbar c^2e\dot A\sigma_x]\psi=0,
\label{Volkov}
\end{split}
\end{equation}
and we only need to remember to check that our solution indeed satisfies the first order Eq. (\ref{Dirac}) at the initial time i.e. $\psi(t=-\infty)=\psi_0$. We can now easily solve Eq. (\ref{Volkov}) with the following ansatz:
\begin{equation}
\psi({\bf r},t)=e^{\frac{i}{\hbar}({\bf p}\cdot{\bf r}-Et)}\phi(u),
\end{equation}
since in the resonant case, the second order Eq. (\ref{Volkov}) actually reduces to a simpler first order Eq.:
\begin{equation}
\dot\phi=\left(i\frac{\dot s}{\hbar}-\frac{1+\sigma_x}{2}\frac{ce\dot A}{E-cp_x+ceA}\right)\phi,
\label{phi}
\end{equation}
where $s(u)$ is given by the Eq. (\ref{s}). We can then easily integrate Eq. (\ref{phi}) to obtain:
\begin{equation}
\phi=e^{\left(i\frac{s-s_0}{\hbar}-\frac{1+\sigma_x}{2}\ln\frac{E-cp_x+ceA}{E-cp_x+ceA_0}\right)}\phi_0,
\end{equation}
where the initial conditions at $u_0=-\infty$ are: $s_0=0$, $A_0=0$, while $\phi_0$ is given by the Eq. (\ref{eigenfunction}). Finally, by using the following Pauli matrix relation:
\begin{equation}
e^{-\frac{1+\sigma_x}{2}\alpha}=\frac{1-\sigma_x}{2}+\frac{1+\sigma_x}{2}e^{-\alpha},
\end{equation}
we obtain the required solution:
\begin{equation}
\psi=e^{iS/\hbar}\left(\frac{1-\sigma_x}{2}+\frac{1+\sigma_x}{2}\frac{E-cp_x}{E-cp_x+ceA}\right)\phi_0,
\label{wavefion}
\end{equation}
where $S$ is the classical action given by the Eq. (\ref{S}) and we avoid the pole by the Landau rule. We can then write the particle probability density as:
\begin{equation}
\begin{split}
&|\psi|^2=e^{-2\Im S/\hbar}\times\\ 
&\phi_0^*\left(\frac{1-\sigma_x}{2}+\frac{1+\sigma_x}{2}\frac{(E-cp_x)^2}{(E-cp_x+ceA)^2+\eta^2}\right)\phi_0,
\end{split}
\label{probability}
\end{equation}
which is shown in the figure \ref{fig1} (a) for the particle that was initially at rest $(p_{x,y}=0)$ and then accelerated in the harmonic field: $A=A_m\sin{k(ct-x)}$, with amplitude $A_m/A_s=3$. One can clearly see the singularity at the point given by the Eq. (\ref{singularity}), signifying the onset of the shockwave. 

Note also the rapid oscillations in the wave function, given by the Eq. (\ref{wavefion}), as we approach the singularity where the phase diverges: $S\rightarrow\infty$. This is in accord with the classical analysis which suggested that the particle gets accelerated to a high energy state. But of course, this can only continue until we reach the limit of our quasi-relativistic approximation, when the actual electron velocity finally exceeds the quasi-limiting velocity. 
One can then view this change in electron dispersion as a perturbation potential which will scatter our quasi-relativistic states. This will result in the reflection of these states with some reduced amplitude, in agreement with the Eq. (\ref{probability}), and confirming that we have chosen the correct sign of the $\eta$. This scattering in a way also represents the mentioned coupling with the environment. Note that for larger quasi-classical cut off energy, we approach more closely the ideal $\eta\rightarrow0^+$ limit, but is would also be interesting to investigate more closely the best way to approximate the finite cut off case with a finite $\eta$.

Let us next show that this singularity leads to energy dissipation, which is also a typical shockwave property \cite{FM, ECM}. Since the electric field has only the longitudinal component: $E_x=-\partial_tA=-c\dot A$, and $|\psi|^2$ is independent of the transverse $y$-coordinate, we can write the dissipated power per transverse unit length as:
\begin{equation}
P/L_y=\int_{-\infty}^{\infty} dx j_xE_x=\int_{-\infty}^{\infty}duj_x(-c\dot A).
\end{equation}
Finally, by using the Dirac current density given by \cite{Jablan20}: $j_x=ec\psi^*\sigma_x\psi$, we get the dissipated power:
\begin{equation}
\begin{split}
&P/L_y=-ec^2\int_{-\infty}^{\infty}du\dot A e^{-2\Im S/\hbar}\times \\
&\phi_0^*\left(\frac{\sigma_x-1}{2}+\frac{1+\sigma_x}{2}\frac{(E-cp_x)^2}{(E-cp_x+ceA)^2+\eta^2}\right)\phi_0.
\label{power}
\end{split}
\end{equation}

Let us now assume that we turn the field off both in the distant past and the future: $A(u=\pm\infty)=0$. Then if are working with weak fields below the shockwave limit (given by the Eq. (\ref{singularity})), there are no singularities in the integral (\ref{power}), $\Im S=0$, and the power simply vanishes: 
\begin{equation}
P\propto\int_{-\infty}^{\infty}du\dot A f(A)=\int_0^0dA f(A)=0.
\label{noloss}
\end{equation}

However, a very different situation occurs in the shockwave regime. Let us say that we hit the singularity of Eq. (\ref{singularity}) $N$ times during the oscillations of the field ($N$ must be even since $A(u=\pm\infty)=0$). Then in between each two consecutive singularities, we get an integral of the type: $P\propto\int_{A_s}^{A_s}dA=0$, which doesn't contribute to the dissipated power. Only parts that do contribute are before the first shock: $P\propto\int_{0}^{A_s}dA$, and after the last shock: $P\propto\int_{A_s}^{0}dAe^{-2\Im S/\hbar}$. We see that we get power dissipation precisely because of this jump in the shockwave probability density $\propto e^{-2\Im S/\hbar}$.

There is only one tiny problem left since the dissipated power given by the Eq. (\ref{power}) actually diverges in the $\eta\rightarrow 0^+$ limit. However, this is just the consequence of the shockwave singularity which means that even the probability density from the Eq. (\ref{probability}) diverges. On the other hand, it should not be too surprising that on resonance we get wave functions which are not normalizable, since a similar thing happens with quasi-discrete levels on resonance (the so-called Gamow states) \cite{QM}. Specifically, to resolve this divergence issue in our case, we need to carefully analyze the $\eta\rightarrow 0^+$ limit. Let us first look at the part of the integral (\ref{power}) from the initial time at $u_0=-\infty$ until the first shock at $u_1$:
\begin{equation}
\begin{split}
&\int_{-\infty}^{u_1}du\dot A\ \frac{(E-cp_x)^2}{(E-cp_x+ceA)^2+\eta^2}=\\
&\int_{0}^{A_s}dA\ \frac{(E-cp_x)^2}{c^2e^2(A-A_s)^2+\eta^2},
\end{split}
\end{equation}
where the lower integral limit is: $A_0=A(-\infty)=0$, and the upper integral limit is positive: $A_s=A(u_1)>0$, in the case of a positive energy state $E>0$ (cf. Eq. (\ref{singularity})). We can then simply extend the lower integral limit to $-\infty$, since the integral is completely determined from the points in the vicinity of the singularity $A_s$ (in the $\eta\rightarrow 0^+$ limit). Similar analysis is also valid on the other side of the shock point $u_1$, only then instead of the integral $\int_{-\infty}^{A_s}dA$ we get the integral $\int_{A_s}^{\infty}dA$ (which has the same value) but with an additional factor $e^{-2\Im S_1/\hbar}$. The same thing happens of course also at the second shock point $u_2$ only there we get a different sign of the integral since the potential $A(u)$ is then descending instead of ascending. We can thus write the main part of the dissipated power (per unit length) as a sum:
\begin{equation}
\begin{split}
P/L_y=-ec^2\sum_{n=1}^N\left(e^{-\frac{2}{\hbar}\Im S_{n-1}}+e^{-\frac{2}{\hbar}\Im S_n}\right)(-1)^{n-1}& \\
\phi_0^*\frac{1+\sigma_x}{2}\phi_0\int_{-\infty}^{A_s}
dA\frac{(E-cp_x)^2}{c^2e^2(A-A_s)^2+\eta^2},&
\end{split}
\end{equation}
while the remaining parts are negligible in the $\eta\rightarrow0^+$ limit. Here $\Im S_n=\Im S(u_n)$, and all the contributions from the middle shocks get canceled so we are only left with the first and the last shock, which determine the dissipated power (as we have already discussed).

We can similarly write the particle probability per transverse unit length as:
\begin{equation}
\begin{split}
\mathcal{P}/L_y=\int_{-\infty}^{\infty}dx|\psi|^2=
\sum_{n=1}^N \left(e^{-\frac{2}{\hbar}\Im S_{n-1}}+e^{-\frac{2}{\hbar}\Im S_n}\right)&\\
\frac{(-1)^{n-1}}{\dot A_n}\phi_0^*\frac{1+\sigma_x}{2}\phi_0\int_{-\infty}^{A_s}
dA\frac{(E-cp_x)^2}{c^2e^2(A-A_s)^2+\eta^2},&
\end{split}
\end{equation}
where we have changed the integration variable into: $dx=-du=-dA/\dot A$. Finally, we obtain a finite expression of the normalized power (in the $\eta\rightarrow0^+$ limit):
\begin{equation}
P/\mathcal{P}=ec^2\frac{e^{-\frac{2}{\hbar}\Im S_N}-1}
{\sum_{n=1}^N \frac{(-1)^{n-1}}{\dot A_n} \left(e^{-\frac{2}{\hbar}\Im S_{n-1}}+e^{-\frac{2}{\hbar}\Im S_n}\right)}.
\end{equation}

Specifically, in the case of the symmetric field where: $\dot A_n=\dot A_1 (-1)^{n-1}$, we get a simple sum of the geometric order which can be simply evaluated to obtain a neat expression:
\begin{equation}
P/\mathcal{P}=ec^2\dot A_1\frac{e^{-\frac{2}{\hbar}\Im S_1}-1}{e^{-\frac{2}{\hbar}\Im S_1}+1},
\label{neat}
\end{equation}
which simplifies further at large fields when: $2\Im S_1/\hbar\ll1$, and we obtain the dissipated power:
\begin{equation}
P/\mathcal{P}\approx\frac{\pi(E^2-c^2p_x^2)}{2\hbar}=P_d.
\end{equation}
Note that we get the correct sign of the energy dissipation ($P_d>0$ for the positive energy states $E>0$) confirming that we have chosen the right sign of $\eta$. Particularly in the case of harmonic field: $A(u)=A_m\sin ku$, we have:
\begin{equation}
\dot A_1=kA_s\sqrt{(A_m/A_s)^2-1},\ A_s=\frac{E-cp_x}{c|e|},
\label{As}
\end{equation}
\begin{equation}
\frac{2}{\hbar}\Im S_1=\frac{A_d/A_s}{\sqrt{(A_m/A_s)^2-1}},
\ \frac{A_d}{A_s}=\pi\frac{E+cp_x}{\hbar ck},
\end{equation}
so that we asymptotically reach the power: $P/\mathcal{P}\rightarrow P_d$, for large fields: $A_m\gg A_d$ (see figure \ref{fig1} (b)).

To give some sense of the scale, we can look at the shockwave threshold given by the potential $A_s$ from the Eq. (\ref{As}). E.g. for an infrared frequency: $\omega=ck=100\ \mbox{THz}$, and electron that was initially at rest $(p_{x,y}=0)$ we need a very low electric field: $E_s=\omega A_s=1\ \mbox{kV/m}$, which can be reduced even further by working at smaller frequencies.
This field is low because of the low mass (i.e. the rest energy $mc^2\approx10\ \mu\mbox{eV}$) of the electron in graphene and can be reduced even further by increasing the electron momentum $p_x$ along the field (cf. Eq. (\ref{As})). The reason for such a strong nonlinear (shockwave) response at such low fields is precisely the resonant nature of the effect. Moreover, while typical resonant transitions between discrete levels yield strong nonlinear response only for discrete frequencies, the velocity resonance (described in this paper) yields a strong nonlinear response over a broadband spectrum.

On the other hand, if one wants to obtain a strong power dissipation $P_d$, it would be best to reduce $p_x$ and increase the electron momentum $p_y$ perpendicular to the field (which gives the electron energy: $E\approx cp_y$), but then one also needs much larger fields. E.g. for some typical electron energies in graphene \cite{Jablan20}: $E=0.1\ \mbox{eV}$, the threshold field is: $E_s=10\ \mbox{MV/m}$, while at fields: $E_m=E_d\approx 5E_s$, we reach the dissipated power maximum: $P_d\approx4\ \mu\mbox{W}$ (case shown in the figure \ref{fig1} (b)). 
Even though this seems like a tiny dissipation for a single particle, it can grow to a considerable amount by using the large electron density in graphene \cite{Jablan20}: $n\approx10^{16}\ \mbox{m}^{-2}$. Namely, the dissipated energy during one oscillation period $(T=2\pi/\omega)$ per unit area is then roughly: $nP_dT\approx3\ \mbox{mJ/m}^2$. To get a more accurate result we can simply integrate dissipated power from the Eq. (\ref{neat}) over the full conduction band with the Fermi energy $E_F=0.1\ \mbox{eV}$, in the field $E_m=E_d=50\ \mbox{MV/m}$, which gives dissipated energy density of $0.8\ \mbox{mJ/m}^2$. This is even larger than the EM energy density \cite{Jablan20}: $\varepsilon_0|E_m|^2/k\approx0.2\ \mbox{mJ/m}^2$, and would result in huge EM damping. Here we have assumed that our longitudinal field is due to some 2D plasmon oscillation which is localized to the order of a wavelength $(\sim1/k)$ in a perpendicular direction. At larger fields $(E_m>E_d)$ energy density grows further but dissipation saturates (see figure \ref{fig1} (b)) so the field damping reduces, while the field is strongly damped at lower fields $(E_m<E_d)$.

The simplest way to test our predictions would be to prepare first the single electron states (e.g. via the Levitov injection \cite{Leviton06, Leviton23}) and then locate the shockwave singularities with the high-resolution microscopy (e.g. using high speed scanning tunneling microscope (STM) \cite{STM23} or scanning near field optical microscope (SNOM) \cite{SNOM25}). 
To prepare the longitudinal field itself, one could use a line of charges moving parallel to the graphene plane or simply use the longitudinal field from say some external plasma oscillations \cite{Jablan20} or a waveguide \cite{ECM}. 

On the other hand, to measure the dissipation in the field one would certainly need to use a group of particle states with some spread in the momentum values $(p_x,p_y)$ to enhance the effect. However, since particles get localized at the shockwave singularities one might worry that Pauli principle would interfere with the many body shockwave properties. This is in fact not so as one can easily show from the Eq. (\ref{singularity}) that states with different initial momentum values $(p_x,p_y)$ get localized at different points in space. However, this is true only as long as we are working within the low-energy quasi-relativistic approximation, while the high-energy corrections would start to scatter our quasi-relativistic states (as we already discussed). 
Pauli principle would then completely block this intraband scattering and would not allow any gain or loss of energy from the field in case of say a full valence band. Of course, Pauli principle still allows interband transitions which will dissipate energy, but these are suppressed unless the photon energy is comparable to the high energy quasi-relativistic cut off (which is $\sim$ eV in graphene \cite{Dirac14}). In any case, it would be interesting to study further corrections to this quasi-classical approximation.  

Note that energy dissipation is actually quite common shockwave property weather one is working with fluid waves \cite{FM} or electromagnetic waves \cite{ECM}. 
However, we need to emphasize that here we described an actual quantum shockwave in the single particle wave function, which is very different from e.g. quantum corrections to the many body mean field shockwave approximations discussed recently in quantum fluids \cite{QShock19, QShock20, QShock25}.
In fact, since this shockwave is a genuinely quantum phenomena one can immediately think about many subtleties of the quantum physics involved. For example, the localization of the particle at the shockwave singularities raises an interesting question involving the quantum measurement problem. Issue that could most precisely studied in the sterile conditions of ultracold atomic gasses where one can as well create quasi-relativistic particle dispersion \cite{Dirac14}.

Also, even though we have focused only on a rather simple property of power dissipation in these shockwave states, it would be interesting to study further e.g. the absorption or emission of light in these states, similar to the study of these effects in the Volkov states \cite{QED}. In fact, since particle gets suddenly localized at the shockwave singularity, one would expect to see strong emission of light with specific profile, which could also be used to detect this shockwave in the far field.

Furthermore, even though we have focused on 2D Dirac electrons in graphene, it is straight forward to show that similar shockwave solutions occur in 3D and 1D Dirac systems. In fact, it would be especially interesting to explore the nature of this shockwave in the 1D systems of Luttinger liquids where particle interactions are particularly strong \cite{Haldane81}.

Before closing we note that the reference \cite{Floquet25} just recently discussed the optical response of Dirac electrons in the Floquet formalism, confirming that there is a strong nonlinear response near the resonance, as was shown earlier in the reference \cite{Jablan20}. Interestingly, reference \cite{Floquet25} does point out the possibility of the shockwave on resonance but does not give any analysis of the resonant (i.e. the shockwave) response. Also, some of these Volkov type solutions in longitudinal EM fields have been recently discussed in the reference \cite{Volkov11} however in a somewhat approximate manner and without any mention of the shockwave solution. Note also that the original Volkov analysis of transverse EM field \cite{QED} does not lead to a shockwave response. Finally note that this quantum shockwave, occurring when EM field moves at the resonance with the limiting particle velocity, is very different from the Cherenkov radiation which occurs when a particle moves faster than the velocity of EM field \cite{ECM}.

In conclusion we studied the resonant interaction of a longitudinal electromagnetic field and a Dirac electron in graphene. We showed that a strong field leads to the quantum shockwave, which is accompanied by energy dissipation, while the electron gets localized at the moving shockwave singularity. Effect that is especially strong in graphene for two reasons: in the single-particle case we get a shockwave at very small fields due to the low electron mass, while in the many-particle case we can get large dissipation by using the large electron density. While the many-particle dissipation can be easily detected due to the rapid decay of the resonant field, the single-particle quantum shockwave can be most easily detected by locating the shockwave singularity with a high-resolution microscopy.

This work was supported by the European Commission and the Croatian Ministry of Science, Education and Sports Co-Financing Agreement No. 291823, Marie Curie FP7-PEOPLE-2011-COFUND NEWFELPRO project GRANQO, and QuantiXLie Centre of Excellence, a project cofinanced by the Croatian Government and European Union through the European Regional Development Fund - the Competitiveness and Cohesion Operational Programme (Grant KK.01.1.1.01.0004).


\begin{thebibliography}{99}

\bibitem{QED}
V. B. Berestetskii, E. M. Lifshitz, and L. P. Pitaevskii, Quantum Electrodynamics, 2nd Edition (Butterworth-Heinemann, Oxford, 1999).

\bibitem{Dirac14}
T. O. Wehling, A. M. Black-Schaffer, and A. V. Balatsky, Dirac materials, Adv. in Phys. {\bf 63}, 1 (2014).

\bibitem{Jablan20}
M. Jablan, Quasiclassical nonlinear plasmon resonance in graphene, Phys. Rev. B {\bf 101}, 085424 (2020).

\bibitem{QM}
L. D. Landau, and E. M. Lifshitz, Quantum Mechanics, 3rd Edition (Butterworth-Heinemann, Amsterdam, 2003).

\bibitem{FM}
L. D. Landau, and E. M. Lifshitz, Fluid mechanics, 2nd Edition (Butterworth-Heinemann, Amsterdam, 2009).

\bibitem{ECM}
L. D. Landau, E. M. Lifshitz, and L. P. Pitaevskii, Electrodynamics of Continuous Media, 2nd Edition (Butterworth-Heinemann, Amsterdam, 2007).

\bibitem{Leviton06}
J. Keeling, I. Klich, and L. S. Levitov, Minimal Excitation States of Electrons in One-Dimensional Wires, Phys. Rev. Lett. {\bf 97}, 116403 (2006).

\bibitem{Leviton23}
A. Assouline, L. Pugliese, H. Chakraborti, S. Lee, L. Bernabeu, M. Jo, K. Watanabe, T. Taniguchi, D. C. Glattli, N. Kumada, H.-S. Sim, F. D. Parmentier, and P. Roulleau, Emission and coherent control of Levitons in graphene, Science {\bf 382}, 1260 (2023).

\bibitem{STM23} K. Liang, L. Bi, Q. Zhu, H. Zhou, and S. Li, Ultrafast Dynamics Revealed with Time-Resolved Scanning Tunneling Microscopy: A Review, ACS Appl. Opt. Mater {\bf 1}, 924 (2023).

\bibitem{SNOM25}
R. Hillenbrand, Y. Abate, M. Liu, X. Chen, and D. N. Basov, Visible-to-THz near-field nanoscopy, Nature Rev. Mater. (2025).

\bibitem{QShock19}
T. Veness, and L. I. Glazman, Fate of quantum shock waves at late times, Phys. Rev. B {\bf 100}, 235125 (2019).

\bibitem{QShock20}
S. A. Simmons, F. A. Bayocboc Jr., J. C. Pillay, D. Colas, I. P. McCulloch, and K. V. Kheruntsyan, What is a Quantum Shock Wave?, Phys. Rev. Lett. {\bf 125}, 180401 (2020). 

\bibitem{QShock25}
A. Urilyon, S. Scopa, G. Del Vecchio, and J. De Nardis, Quantum fluctuating theory for one-dimensional shock waves {\bf 111}, 045401 (2025).

\bibitem{Haldane81}
F. D. M. Haldane, 'Luttinger liquid theory' of one-dimensional quantum fluids. I. Properties of the Luttinger model and their extensions to the general 1D interaction spinless Fermi gas, J. Phys. C {\bf 14}, 2585 (1981).

\bibitem{Floquet25}
T. Oka, Shockwave-Enhanced Floquet Engineering in Relativistic Quasiparticles, arxiv:2407.21458 (2025).

\bibitem{Volkov11}
J. T. Mendonca, A. Serbeto, Volkov solutions for relativistic quantum plasmas, Phys. Rev. E {\bf 83}, 026406 (2011).


\end{thebibliography}
\end{document}